\documentclass[doublecol]{epl2} 

\title{Non-Hermitian quantum dynamics and entanglement of coupled nonlinear resonators}
\shorttitle{Non-Hermitian quantum dynamics and entanglement of coupled resonators} 

\author{Evren~Karakaya\inst{1} \and Ferdi~Altintas\inst{2} \and Kaan~G\"{u}ven\inst{1} 
\and \"{O}zg\"{u}r~E.~M\"{u}stecapl{\i}o\u{g}lu\inst{1}}
\shortauthor{E.~Karakaya \etal}

\institute{
  \inst{1} Department of Physics, Ko\c{c} University, Sar{\i}yer, \.Istanbul, 34450, Turkey\\
  \inst{2} Department of Physics, Abant Izzet Baysal University, Bolu, 14280, Turkey 
}

\abstract{We consider a generalization of recently proposed non-Hermitian model for resonant cavities coupled by a chiral mirror by taking into account number non-conservation and nonlinear interactions. We analyze non-Hermitian  quantum dynamics of populations and entanglement of the cavity modes.
We find that an interplay of initial coherence and non-Hermitian coupling leads to a counterintuitive population transfer. While an initially coherent cavity mode is depleted, the other empty cavity can be populated more or less than the initially filled one. Moreover, presence of nonlinearity yields population collapse and revival as well as bipartite entanglement of the cavity modes. In addition to coupled cavities, we point out that similar models can be found in $\mathcal{PT}$ symmetric Bose-Hubbard dimers of Bose-Einstein condensates or in coupled soliton-plasmon waveguides. We specifically illustrate quantum dynamics of populations and entanglement in a heuristic model that we propose for a soliton-plasmon system with soliton amplitude dependent asymmetric interaction. Degree of asymmetry, nonlinearity and coherence are examined to control plasmon excitations and soliton-plasmon entanglement. Relations to $\mathcal{PT}$ symmetric lasers and Jahn-Teller systems are pointed out.
}

\begin{document}

\maketitle

\section{Introduction}

Recently, an intriguing quantum optical model for resonant cavities coupled by a chiral mirror has been proposed~\cite{santos12}. The chiral mirror is a planar metamaterial array of asymmetric split rings, through which transmission of circularly polarized electromagnetic waves becomes different in the opposite direction without violating Lorentz reciprocity principle~\cite{fedotov06}. The transmission matrix of such a mirror is a two dimensional non-Hermitian matrix. The proposed quantum optical system is then described by a non-Hermitian Hamiltonian. As a special class of non-Hermitian systems, parity-time ($\mathcal{PT}$) symmetric systems have been attracted much attention~\cite{boettcher98,bender07,mostafazadeh02}. More recently, it is shown that hybridized metamaterials can simulate effectively spontaneous symmetry breaking in  $\mathcal{PT}$- symmetric non-Hermitian quantum systems~\cite{kang13}. Nonreciprocal light transmission in $\mathcal{PT}$- symmetry broken phase using whispering-gallery microcavities has been observed very recently~\cite{peng13}.

Non-Hermitian interactions are also reported for Bose-Einstein condensates~\cite{graefe08,dast13}, optical lattices~\cite{regensburger13}, waveguides~\cite{klaiman08,bludov13,honghua14} 
and soliton-plasmon systems~\cite{milian12,ferrando13}. Soliton-plasmon system has unique properties such as nonlinear coupling in addition to local nonlinearity due to soliton and can be examined from a pure quantum perspective in terms of generalized models of coupled cavities~\cite{santos12} or Bose-Einstein condensates~\cite{graefe08}. Effects of local nonlinearities on systems with $\mathcal{PT}$ symmetry have been subject to recent attention~\cite{lumer13,miroshnichenko11}. Moreover, chiral mirror coupled cavities or soliton-plasmon systems we examine here have non-local non-Hermiticity in the coupling in contrast to typical systems considered in literature with local non-Hermiticity associted with balanced gain and loss.

The non-Hermitian quantum model for resonant cavities coupled by a chiral mirror uses a number conserving Heisenberg approach to determine the quantum dynamics~\cite{santos12}. In the present letter, we use a more general von Neumann approach~\cite{sergi13,graefeKorsch08,dattoli90,brody12} which allows for violation of number conservation due to the non-Hermitian nature of the system. These two approaches are not equivalent to each other and von Neumann approach is proper treatment of non-conservative systems with the non-Hermitian models. We determine the population dynamics of the resonator modes when one resonator is in a coherent state and the other is empty. Strong excitations of an initially empty resonator mode is found to be possible by a weakly populated coherent one depending on the asymmetry of their coupling. Curious role of coherence on particle non-conservation is revealed. In addition we characterize bipartite quantum entanglement of the resonators by using von Neumann entropy~\cite{siva09} and as expected initially unentagled state remains factorized with zero entropy. This is typical as one cannot create entanglement by mixing
coherent states~\cite{kim02,xiang02}.
We then propose generalized models where resonator and their coupling could be nonlinear. 
In this case, quantum entanglement between the modes emerges. 
We emphasize that the proposed nonlinear models are physically different from the coupled linear resonator model.

The specific purpose of the present letter is to propose models which are experimentally achievable in soliton-plasmon and in Bose-Hubbard dimers as well as in nonlinear coupled cavities, and to reveal the interplay of coherence, local and non-local nonlinearities, and asymmetric non-local non-Hermitian coupling on quantum entanglement and on number non-conserving nonlinear quantum dynamics.
Our models are of broad interest to the subjects of cavity QED, Bose-Einstein condensates, coupled waveguides, optical solitons and quantum plasmonics. We point out a close relation to Jahn-Teller physics~\cite{wu12} and $\mathcal{PT}$ symmetric lasers and anti-lasers as well~\cite{chong10,longhi10,liertzer12,yoo11}. 

From more fundamental perspective, 
one can learn more about the $\mathcal{PT}$ symmetry in such systems by further studies of the spectral properties. 
In contrast to typical classical electrodynamical
$\mathcal{PT}$ symmetric systems we have pure quantum models which could be used to examine quantum correlations and $\mathcal{PT}$ symmetry relations. Our von Neumann approach can be extended to open system conditions, which can be implemented by the coupled cavity and soliton models, to examine $\mathcal{PT}$ symmetry effects in open system. In contrast to typical local $\mathcal{PT}$ symmetric terms, our models have non-local non-reciprocal coupling as a different class of $\mathcal{PT}$ symmetry and hence they can bring new perspectives to nonlinearity, non-locality and broken $\mathcal{PT}$ symmetry relations.
\section{Non-Hermitian dynamics and Entanglement} 
Before investigating the mean-field dynamics and the entanglement between coupled field modes in two experimentally feasible quantum models which are represented by non-Hermitian Hamiltonians, we give 
a brief introduction to the non-Hermitian dynamics and the entanglement measure. 

The general formulation for the time evolution of quantum systems under non-Hermitian Hamiltonians can be found in~\cite{sergi13,graefeKorsch08,dattoli90,brody12}. A non-Hermitian Hamiltonian operator can be partitioned into Hermitian and anti-Hermitian parts:
\begin{eqnarray}\label{ham}
H=H_{+}+H_{-},
\end{eqnarray}
where $H_{\pm}=\frac{1}{2}\left(H\pm H^{\dagger}\right)$, $H_{\pm}=\pm H_{\pm}^{\dagger}$, and $A^{\dagger}$ denotes the Hermitian conjugate of $A$. The non-Hermitian Schr\"{o}dinger equation can be written as 
(Here and in the rest of the paper we take $\hbar = 1$.)
\begin{eqnarray}\label{se}
\frac{\partial}{\partial t}\left|\Psi(t)\right\rangle=-i\left(H_{+}+H_{-}\right)\left|\Psi(t)\right\rangle .
\end{eqnarray}
Introducing the density matrix $\rho(t)=\left|\Psi(t)\right\rangle\left\langle \Psi(t)\right|$ leads to the von Neumann type master equation in the following form~\cite{sergi13}:
\begin{eqnarray}\label{me}
\frac{\partial}{\partial t}\rho(t)=-i[H_{+},\rho(t)]-i\{H_{-},\rho(t)\},
\end{eqnarray}
where $[\ldots]$ and $\{\ldots\}$ stand for commutator and anti-commutator, respectively. Here we assure that  Eq.~(\ref{me}) also holds for mixed states~\cite{sergi13}. Since the dynamics governed by~(\ref{se}) is not unitary, the trace of the density operator may not be conserved. Therefore, we introduce a normalized density operator,
\begin{eqnarray}\label{ndm}
\rho^{'}(t)=\rho(t)/Tr(\rho(t)),
\end{eqnarray}
then the time evolution of the quantum averages of the observables can be calculated through the formula
\begin{eqnarray}\label{oa}
\left\langle A\right\rangle_t=Tr\left(A\rho(t)\right)/Tr(\rho(t)).
\end{eqnarray}

Entangled states are essential resource for many applications of quantum information and computation protocols. It would be very desirable to study the possibility of detecting entanglement between the coupled modes~\cite{siva09}. In our study, we consider an initial pure product state of the field modes and have carefully checked that the purity of the considered initial state is conserved during the dynamics~(\ref{me}) in accordance with the fact that an initial pure state remains pure under non-Hermitian dynamics~\cite{brody12}.
The entanglement between coupled modes in a pure state can be quantified by the von Neumann entropy 
\begin{eqnarray}\label{entan}
S=-Tr\left(\rho_{i}^{'}\log_2\rho_{i}^{'}\right),
\end{eqnarray}
where $\rho_{i}^{'}=Tr_j\rho^{'}$ is the reduced density matrix of the mode $i$ obtained by taking a partial trace over the remaining mode $j$. It is zero for separable states and reaches the maximum value of $\log_2N$ for a $N$ dimensional Hilbert space. Non-zero values of the entropy indicate that the modes are in an entangled state.

We apply the formalism described above to systematically examine the entanglement and population dynamics in a generic resonant two-mode model
\begin{eqnarray}
H=\omega_0(a^\dag a+b^\dag b)+ua^\dag a^\dag a a + g_{AB}ab^\dag + g_{BA} f(a^\dag a) a^\dag b,
\end{eqnarray}
where $a\; (a^\dag) $ and $b\; (b^\dagger)$ are the lowering (raising) operators of the bosonic cavity field 
modes A and B, respectively, which obey the Weyl-Heisenberg $h_3$ algebra $[a,a^{\dagger}]=[b,b^{\dagger}]=1$; 
$\omega_0$ is the resonance frequency; $u$ is the parameter of local nonlinearity, while $f(a^\dag a)$ is an intensity dependent deformation function describing non-local nonlinearity. 
Non-reciprocal coupling of the modes is described by the coefficients $g_{AB}$ and $g_{BA}$, which are assumed to be real numbers for simplicity. More
local and non-local sources for nonlinearity could be introduced but this model has optimal number of parameters to capture the essential physics and
relevant to experimental systems. We first consider linear model with 
$u=0$ and $f(a^\dag a)=1$ which corresponds to chiral mirror coupled
optical cavities~\cite{santos12}. After that we discuss the case of local nonlinearity {\it per se} with $f(a^\dag a)=1$ and point out that the model translates to a $\mathcal{PT}$ Bose-Hubbard dimer~\cite{honghua14}. Finally we include the nonlocal nonlinearity by taking $f(a^\dag a)=\sqrt{a^\dag a}$ to make the model relevant to asymmetric amplitude dependent coupled soliton-plasmon system~\cite{milian12}.
\section{Resonant cavities coupled by a chiral mirror}
We first reconsider the non-Hermitian quantum model that describes the behavior of a cavity mode in two resonant cavities coupled through a 2-D chiral mirror~\cite{santos12}. The mode in each cavity can be described by independent quantum oscillators. The Hamiltonian of the system in the weak coupling scenario can be written as
\begin{eqnarray}\label{hm1}
H=H_0+H_I,
\end{eqnarray}
where
\begin{eqnarray}\label{h01}
H_0=\omega_0 a^{\dagger}a+\omega_0 b^{\dagger}b
\end{eqnarray}
is the free Hamiltonian of the oscillators having the transition frequencies $\omega_0$, and
\begin{eqnarray}\label{hi1}
H_I=g_{AB}ab^{\dagger}+g_{BA}a^{\dagger}b
\end{eqnarray}
is the interaction Hamiltonian which describes a non-reciprocal coupling between the oscillators if $g_{AB}\neq g_{BA}$. 

The analytical non-Hermitian dynamics of the mean excitation number in cavity A and B for the considered model have been investigated in Ref.~\cite{santos12} by solving the standard Heisenberg equation of motion in an {\it ad hoc} manner. The exchange of a photon from one cavity to the another through the 2-D chiral mirror has been found. Also, this formalism has been found to preserve the total excitation number, since the total number operator $a^{\dagger}a+b^{\dagger}b$ commutes with the Hamiltonian (\ref{hm1}). On the other hand, the standard Heisenberg equation is based on the Hamiltonian being Hermitian, so it is not expected to capture more general and subtle features of non-Hermitian dynamics. Therefore, our aim in this section is to reexamine the mean-field dynamics by using the formalism described above. 
\begin{figure}[!ht]
\onefigure[width=0.35\textwidth]{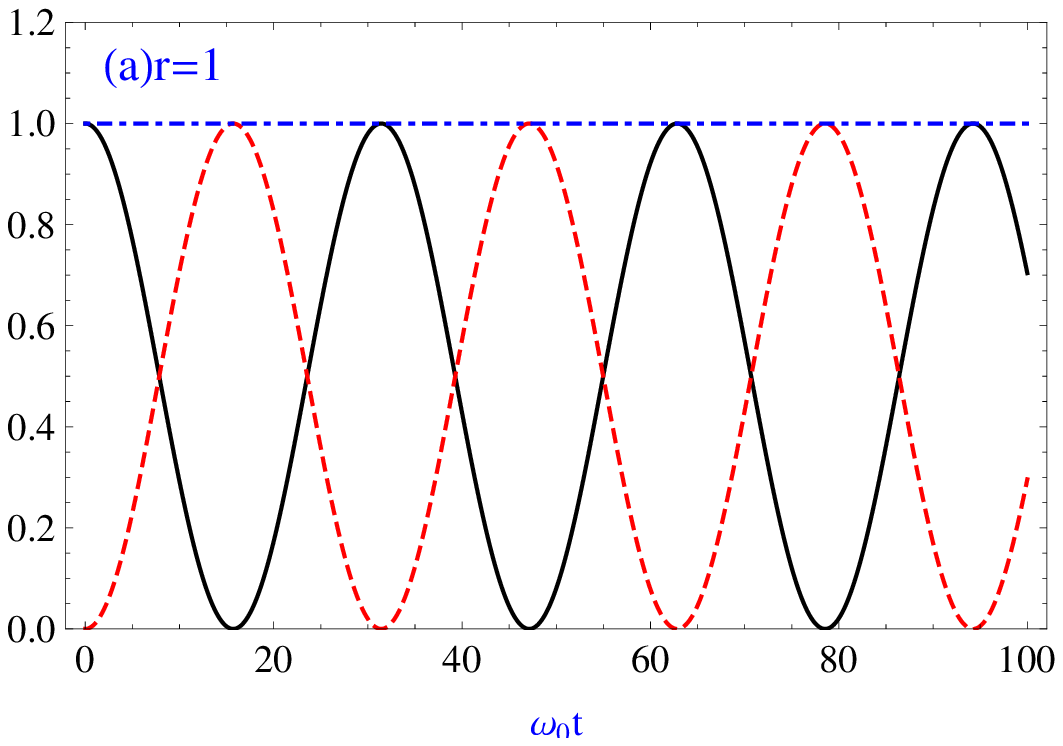}
\onefigure[width=0.35\textwidth]{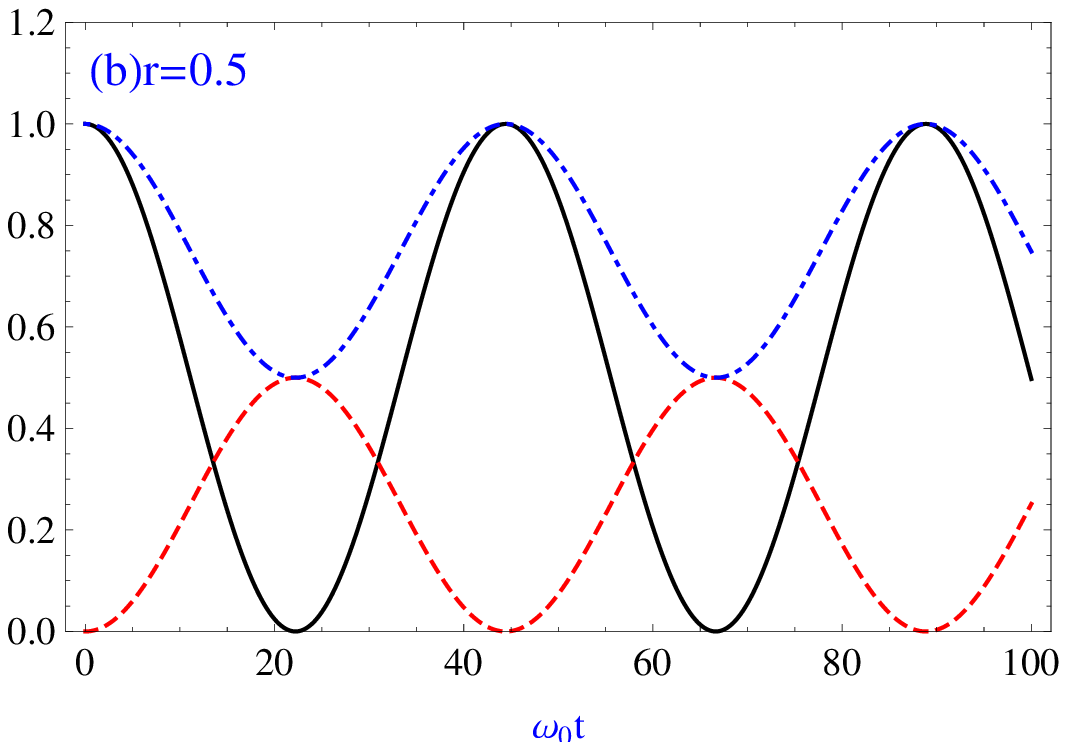}
\onefigure[width=0.35\textwidth]{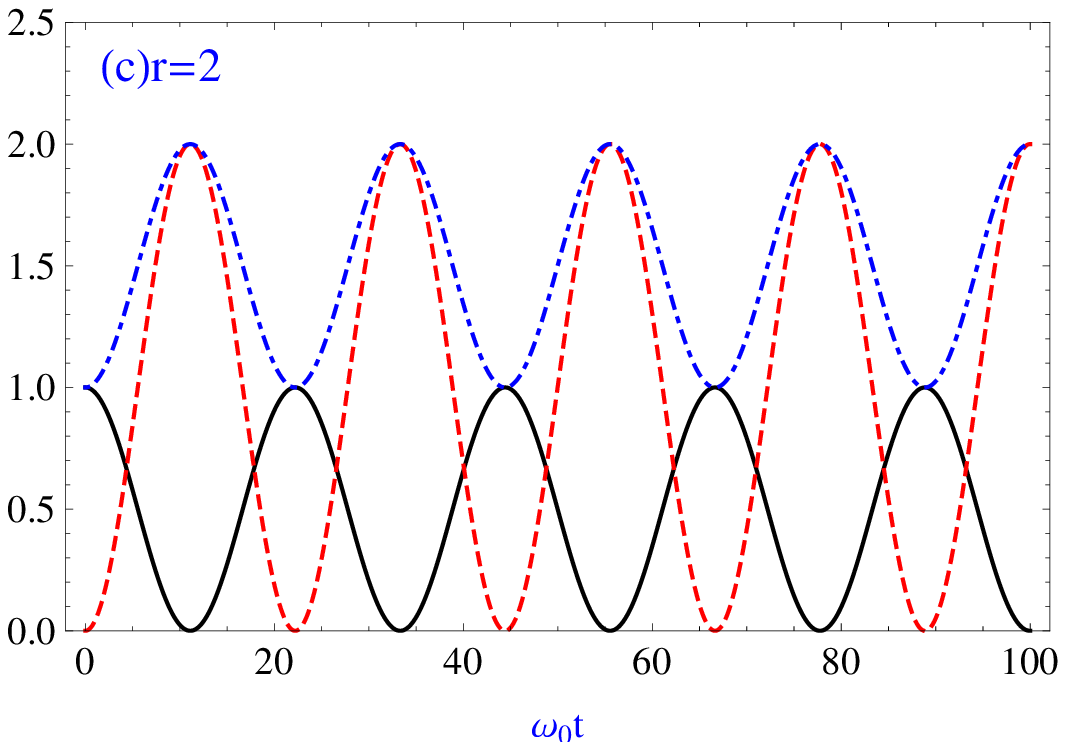}
\caption{$\left\langle a^{\dagger}a\right\rangle_t$~(solid line), $\left\langle b^{\dagger}b\right\rangle_t$~(dashed line) and $\left\langle a^{\dagger}a+b^{\dagger}b\right\rangle_t$~(dot-dashed line) versus $\omega_0 t$ for the initial state $\left|\Psi(0)\right\rangle=\left|\alpha_A\right\rangle\left|0_B\right\rangle$, $\alpha_A=1$, $g_{AB}=g.r$, $g_{BA}=g$, $g=0.1\omega_0$, $r=1$ (a), $r=0.5$ (b) and $r=2$ (c).}
\label{fig1}
\end{figure}

We first derive the Heisenberg equation for the non-Hermitian Hamiltonian, and show that the expectation value of the total number operator may not be constant. 
The time evolution of the expectation value of an observable through the normalized state can be written as: $\left\langle A \right\rangle_t=\left\langle \Psi(t) \right| A \left| \Psi(t)\right\rangle/\langle \Psi(t)|\Psi(t)\rangle$. By using Eq.~(\ref{se}), the generalized Heisenberg equation for the expectation value of an operator can be written as~\cite{graefeKorsch08}:
\begin{eqnarray}\label{ghe}
\frac{\partial}{\partial t}\left\langle A \right\rangle_t=-i\left\langle [A,H_+] \right\rangle_t-i\left\langle \{A,H_-\} \right\rangle_t+2i\left\langle A \right\rangle_t \left\langle H_- \right\rangle_t.
\end{eqnarray}
It is simple to show that the expectation value of the total excitation number evolves as:
\begin{eqnarray}\label{noe}
\frac{\partial}{\partial t}\left\langle a^{\dagger}a+b^{\dagger}b \right\rangle_t&=&i[g_{AB}-g_{BA}][\left\langle(a^{\dagger})^2ab-a^{\dagger}a^2b^{\dagger}\right\rangle_t\nonumber\\
&+&\left\langle a^{\dagger}b-ab^{\dagger}\right\rangle_t+\left\langle b^{\dagger}b^2a^{\dagger}-(b^{\dagger})^2ba\right\rangle_t\nonumber\\
&+&\left\langle a^{\dagger}a+b^{\dagger}b\right\rangle_t\left\langle ab^{\dagger}-a^{\dagger}b\right\rangle_t],
\end{eqnarray}
which can be different than zero if $g_{AB}\neq g_{BA}$. Strictly, the time evolution in Eq.~(\ref{noe}) depends on the initial state, as well. 

In Fig.~\ref{fig1}, we have plotted the time evolution of the mean excitation number in cavity A, $\left\langle a^{\dagger}a\right\rangle_t$, in cavity B, $\left\langle b^{\dagger}b\right\rangle_t$, and the total excitation number, $\left\langle a^{\dagger}a+b^{\dagger}b\right\rangle_t$, by numerically solving 
Eq.~(\ref{me}) for $\rho(t)$ and by employing Eq.~(\ref{oa}), for an initial state where the mode A in the coherent state and the mode B in its vacuum, $\left|\Psi(0)\right\rangle=\left|\alpha_A\right\rangle\left|0_B\right\rangle$, for the parameters $g_{AB}=g.r$, $g_{BA}=g$, $r=1$ (Hermitian case) and $r=0.5,2$ (non-Hermitian case). For the Hermitian case (Fig.~1(a)), the total excitation number is conserved in the system and a single photon can be exchanged through the mirror, as found in Ref.~\cite{santos12}. For the non-Hermitian dynamics, the total excitation number is not conserved. 

The non-conservation is an interplay of coherence and non-Hermitian interaction. We numerically verified that an initial Fock state preserves the total excitation number and leads to similar behavior as in Fig.~1(a). 
If $g_{AB}<g_{BA}$, the creation rate of initially empty mode B by initially coherent mode A is smaller than its destruction, then mode B is weakly excited. 
Even if the mode A is completely depleted, its population can only be partially transferred to mode B as can be seen in Fig.~1(b). In this case non-Hermiticity and initial coherence act as a decoherence and population loss channel. On the other hand, if 
$g_{AB}>g_{BA}$, an amplification of the total excitation is found as in Fig.~1(c). In this case,
weak amplitude Fock number states in the coherent state  of mode A can contribute to excitation of the mode B  due to asymmetric interaction in favor of mode B. As a result mode B can be excited with higher amplitudes then the exciting field. 

Our model can be related to the Jahn-Teller system in classical limit~\cite{wu12}. The states of that system is always a coherent state and there is always number non-conservation. In our quantum system, Fock number states lead to number conservation and the significance of initial coherence is revealed. We emphasize that the photons are not generated or destroyed by the mirror but they are 
already present in the initial coherent state. Number non-conservation is just a shift of the mean excitation number due 
to non-reciprocal photon exchange between the cavities which allows for 
unbalanced growth of less fortunate excitation manifolds relative to the initial average.

Number non-conserving dynamics is related to $\mathcal{PT}$ symmetric lasers and anti-lasers as well~\cite{chong10,longhi10,liertzer12,yoo11}. To see this explicitly we can introduce the operators of non-local modes $c=(a+\mathrm{i}b)/\sqrt{2}$ and $d=(a-\mathrm{i}b)/\sqrt{2}$ for which the non-Hermitian part of the model $H_-=(g_{AB}-g_{BA})(ab^\dag-ba^\dag)/2$ becomes diagonal $H_-= \mathrm{i}\Gamma(c^\dag c-d^\dag d)$ with $\Gamma=(g_{AB}-g_{BA})/2$. The Hermitian part remains Hermitian as $H_+=\omega_0N+\mathrm{i}G(c^\dag d-d^\dag c)$ with $N=c^\dag c+d^\dag d$ and $G=(g_{AB}+g_{BA})/2$. Non-local non-reciprocal coupling of local modes then effectively describes local gain and loss on the non-local modes. In the single excitation manifolds, the Hamiltonian blocks of the local and nonlocal modes $H_{\mathrm{l}}^{(1)}$ and $H_{\mathrm{nl}}^{(1)}$ are respectively given by
\begin{eqnarray}
 H_{\mathrm{l}}^{(1)}&=&\left(\begin{array}{cc}
 				\omega_0 & g_{AB} \\
 				g_{BA} & \omega_0 \end{array} \right),\\
 H_{\mathrm{nl}}^{(1)}&=&\left(\begin{array}{cc}
 				\omega_0+\mathrm{i}\Gamma & \mathrm{i}G \\
 				-\mathrm{i}G & \omega_0-\mathrm{i}\Gamma \end{array} \right).
\end{eqnarray}
Effective model with $\omega_0\pm i\Gamma$ energies of non-local modes are similar to the scenarios proposed to generate $\mathcal{PT}$ lasers using a medium with spatially alternating gain and loss regions. Using chiral mirrors could be an efficient and effective way to design $\mathcal{PT}$ symmetric lasers and anti-lasers.

We have also analyzed whether or not the modes become entangled in the processes shown in Fig.~1 via the von Neumann entropy~(\ref{entan}) and by using the reduced density matrices of the modes $A$ and $B$. Our results demonstrate that the entanglement entropy is always zero and hence the states of the modes remain factorized. This is expected as a beam splitter type bilinear mixing interaction cannot be used to create entanglement out of initial coherent states~\cite{kim02,xiang02}. If at least one of the cavities contains a local nonlinearity (for example of Kerr type $ua^\dag a^\dag a a$), then the cavity modes is found to be entangled. Furthermore the population oscillations exhibit collapse and revival dynamics. These effects are due to non-Hermitian and nonlinear interactions and in principle could be realized not only in coupled cavities but in other systems as well. We first show similarity to Bose-Hubbard dimer of a Bose-Einstein condensate model and then propose another system of coupled 
soliton-plasmon waveguides below to illustrate these effects.

\section{Bose-Hubbard dimer of a Bose-Einstein condensate}
Atomic Bose-Einstein condensate trapped in a double well potential possesses local Kerr type 
nonlinearity due to atomic collisions at the trap sites. Generalization of the two-mode condensate model to a $\mathcal{PT}$ symmetric
non-Hermitian model $H_{\mathrm{BEC}}$ has been discussed in the literature~\cite{graefe08} in the form
\begin{eqnarray}
H_{\mathrm{BEC}}=-\mathrm{i}2\gamma L_z+2vL_x+2cL_z^2,
\end{eqnarray}
where $\gamma,v,c$ are constants, and $L_+=L_x+\mathrm{i}L_y=a^\dag b, L_-=L_+^\dag$ and $L_z=(a^\dag a-b^\dag b)/2$ are the Schwinger-boson spin representations obeying the $su(2)$ spin algebra. 

The case of asymmetric coupling of the resonators can be translated to such a model and becomes 
\begin{eqnarray}
H=\omega_0N+g_xL_x-\mathrm{i}g_yL_y,
\end{eqnarray}
where $g_x=g_{AB}+g_{BA}, g_y=g_{AB}-g_{BA}$ and $N=a^\dag a+b^\dag b$. Similar model has been discussed very recently~\cite{honghua14}. Nonlinear spin term $2cL_z^2$ can be generated by considering Kerr type nonlinear cavities with additional $N, N^2$ dependent terms. Such terms can be significant due to possible violation of number conservation. For initial conditions which are number conserving, or with small number fluctations, the similarity of the two systems can be improved. We conclude that nonlinear cavities coupled by chiral mirrors can effectively simulate the non-Hermitian Bose-Hubbard dimer of Bose-Condensate in a double well. Below we envision a more general scenario that allows us to include non-local nonlineraties by proposing a heuristic model of soliton-plasmon system.

\section{Non-linear soliton-plasmon interaction}
The second non-Hermitian model that we propose in the present study describes a non-linear quantum interaction between optical soliton photons in a Kerr medium and surface plasmons in a metal through a dielectric layer~\cite{milian12,ferrando13}. One particular advantage of this model is that the nonlinear coupling can be larger and more tunable than the local nonlinearity. 
The Hamiltonian of the system can be written as
\begin{eqnarray}\label{hm2}
H=\omega_0 a^{\dagger}a+u a^{\dagger}a^{\dagger}aa+\omega_0 b^{\dagger}b+g_{AB}ab^{\dagger}+g_{BA}\sqrt{n_A}a^{\dagger}b,
\end{eqnarray}
where $a$ ($a^{\dagger}$) and $b$ ($b^{\dagger}$) are the field operators for the soliton photons and for the surface plasmons, respectively, and $u$ is the Kerr interaction strength.  Here $g_{BA}\sqrt{n_A}$ with $n_A=a^{\dagger}a$ is the nonlinear operator-type interaction strength and describes the weak soliton amplitude. Without such a nonlinear interaction, the model describes linear and nonlinear cavities coupled through a chiral mirror. One should note that the soliton-plasmon Hamiltonian is always non-Hermitian, even in the case $g_{AB}=g_{BA}$.
\begin{figure}[!ht]
\onefigure[width=0.35\textwidth]{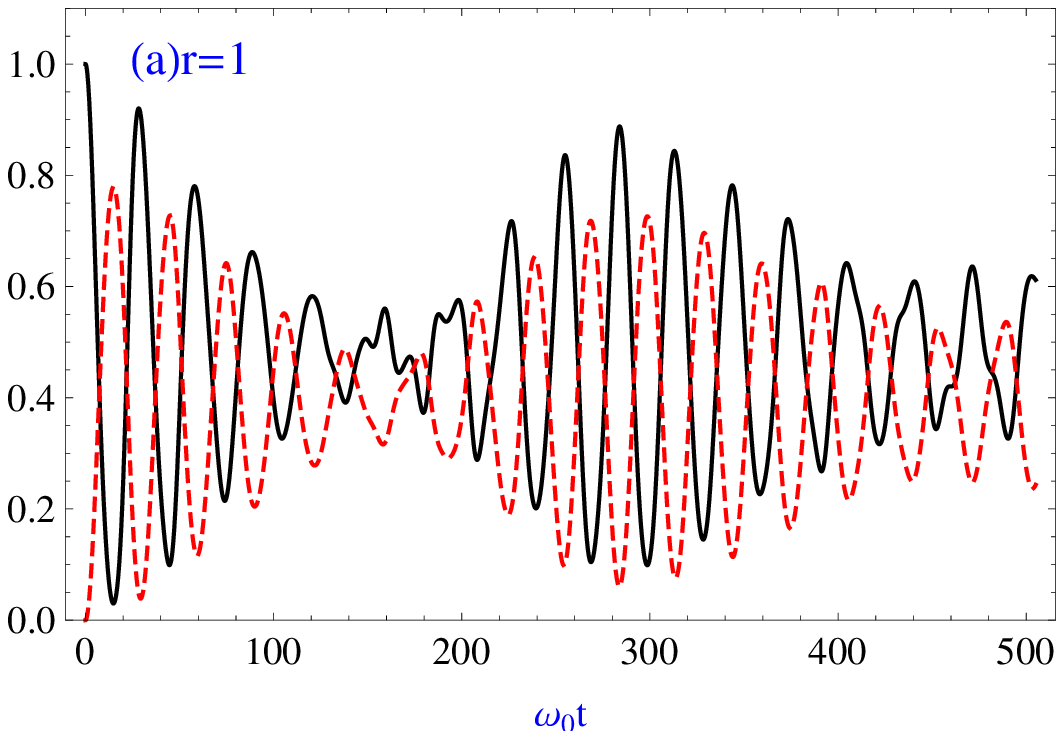}
\onefigure[width=0.35\textwidth]{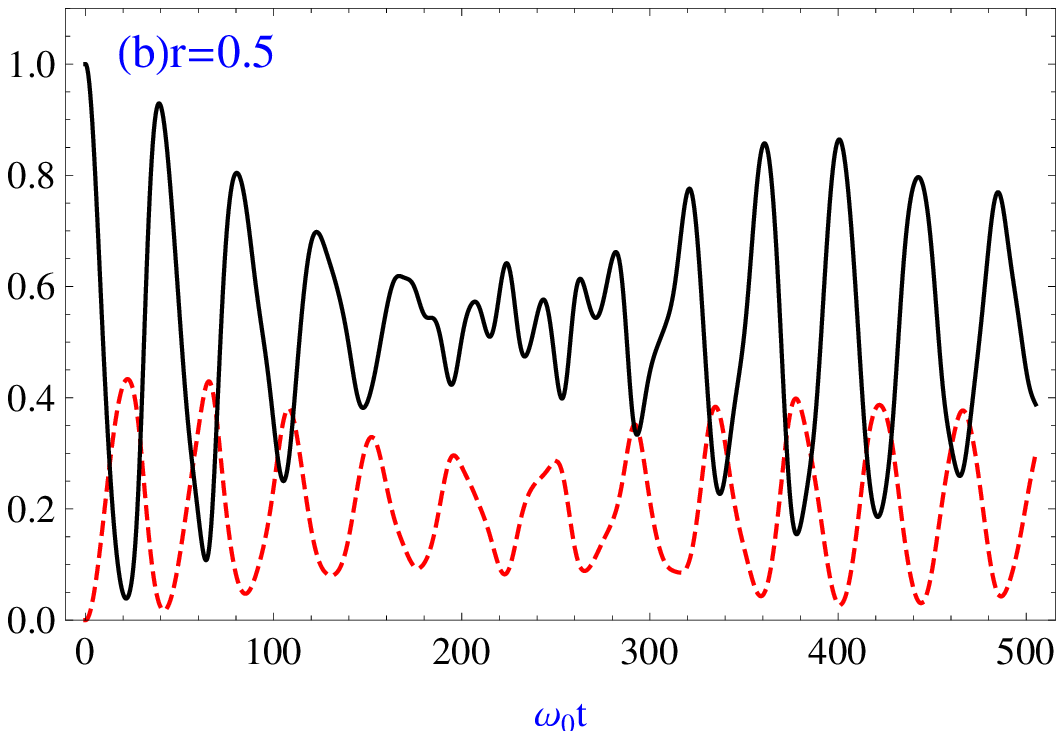}
\onefigure[width=0.35\textwidth]{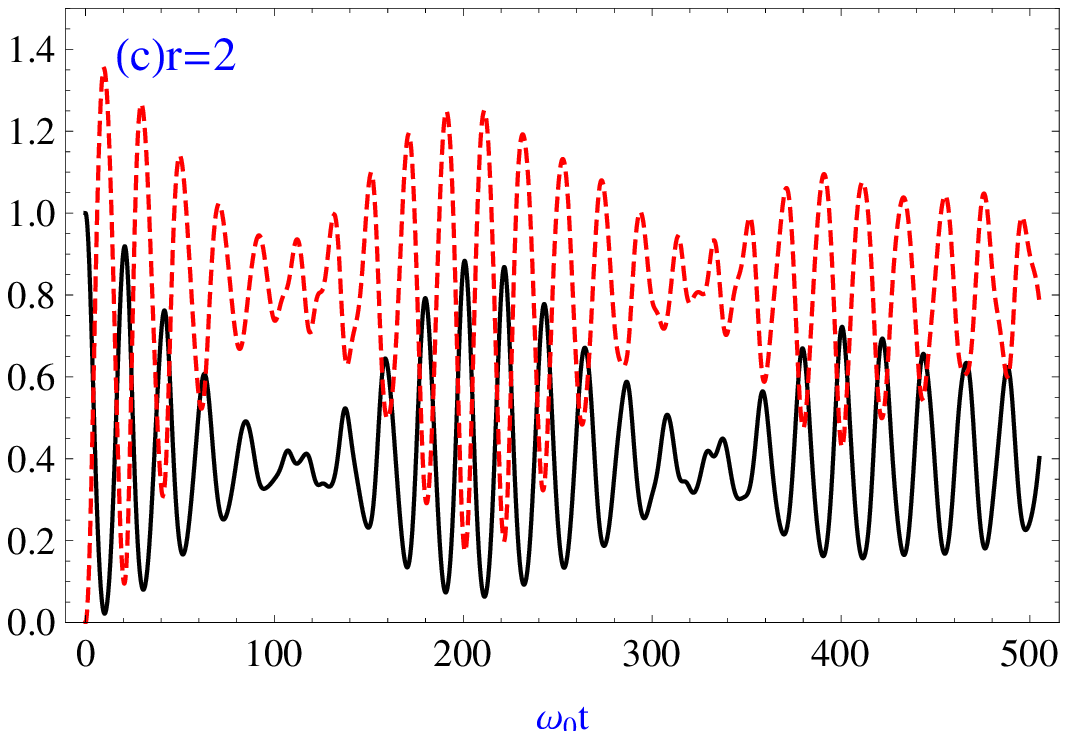}
\caption{$\left\langle a^{\dagger}a\right\rangle_t$~(solid line) and $\left\langle b^{\dagger}b\right\rangle_t$~(dashed line) versus $\omega_0 t$ for the initial state $\left|\Psi(0)\right\rangle=\left|\alpha_A\right\rangle\left|0_B\right\rangle$, $\alpha_A=1$, $g_{AB}=g.r$, $g_{BA}=g$, $g=0.1\omega_0$, $u=-0.01\omega_0$, $r=1$ (a), $r=0.5$ (b) and $r=2$ (c).}
\label{fig2}
\end{figure}

The time dependence of the average soliton photons and surface plasmons number have been plotted in Fig.~2 for the initial state $\left|\Psi(0)\right\rangle=\left|\alpha_A\right\rangle\left|0_B\right\rangle$, where $\left|\alpha\right\rangle$ is the coherent state, $u=-0.01\omega_0$, $g_{AB}=g.r$, $g_{BA}=g$, $g=0.1\omega_0$ and $r=0.5,1,2$. Here we have introduced a small Kerr nonlinearity, so that the system remains almost in soliplasmon resonance. Due to the nature of initial coherent state and the nonlinear interactions, the dynamics of the average excitation numbers exhibit collapse and revival phenomena. Nonlinearity in either local ($u$) or non-local ($g_{BA}$) interaction is sufficient for the collapse-revival effect; though local nonlinearity {\it per se} yields longer time for collapse and revivals. 

The total excitation number in none of the cases is conserved due to the non-Hermitian dynamics. For the cases $r=1$ (Fig.~2(a)) and $r=0.5$ (Fig.~2(b)), the total excitation oscillates below the initial value. For the case $r=2$ (Fig.~2(c)), the total excitation number exceeds its initial value. There is no complete population transfer in soliton photons and surface plasmons, except the $r=2$ case, where time average value of the mean number of plasmon excitations is larger than the time average of the soliton population. In other words, a weak coherent soliton excites strong plasmon, populated more than the initial soliton mode. This happens when $g_{AB}>g_{BA}$ or when transfer rate from soliton to plasmon is greater than the one from plasmon to soliton and the soliton is in the coherent state initially. In the coherent state, weak amplitude Fock states with large photon numbers find chance to contribute to plasmon excitation due to the higher unbalanced transfer rate from soliton to plasmon mode. If we consider initially Fock state for the soliton mode this effect disappears and the number conservation cannot be violated. These conclusions are consistent with the case of coupled cavities by a chiral mirror. 
\begin{figure}[!ht]
\onefigure[width=0.4\textwidth]{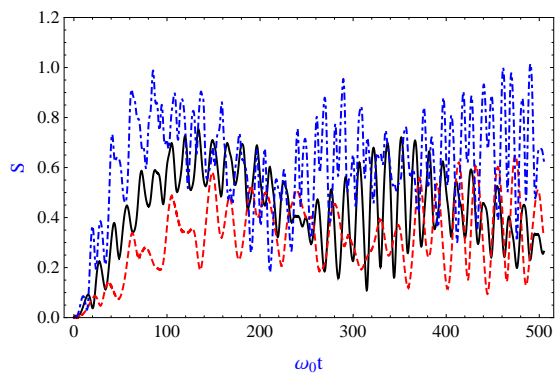}
\caption{Entropy versus $\omega_0 t$ for the initial state $\left|\Psi(0)\right\rangle=\left|\alpha_A=1\right\rangle\left|0_B\right\rangle$, $g_{AB}=g.r$, $g_{BA}=g$, $g=0.1\omega_0$, $u=-0.01\omega_0$, $r=1$ (black, solid), $r=0.5$ (red, dashed) and $r=2$ (blue, dotdashed).}
\label{fig3}
\end{figure}

In Fig.~3, we have investigated the dynamics of entanglement between coupled modes by means of entropy, $S(\rho_A)$, for the same parameters and initial state in Fig.~2. The initial separable soliton photons and surface plasmons become entangled in time. They never become separable in the considered time domain. Contrary to the previous discussed non-Hermitian model (\ref{hm1}), the non-linearities in Eq.~(\ref{hm2}) are found to lead to the bipartite entanglement of the modes. The entropy increases in the collapse region of the populations and takes its largest values around the collapse time. The largest entropy value is found for the case of strong plasmon excitation ($r=2$). The maximum entanglement corresponds to maximum entropy $S_{\mathrm{max}}=\log_2M$ for a Hilbert space dimension of $M$.
We take $M=10$ for each soliton and plasmon Fock space. We conclude that the generated bipartite entanglement is not maximal. As in the case of collapse-revival effect, though any source of nonlinearity suffices for entanglement, local nonlinearity {\it per se} would yield longer time to generate largest entanglement.

\section{Conclusions}
We considered effects of local nonlinearity and amplitude dependent, asymmetric interactions on the non-Hermitian quantum dynamics and entanglement of a two resonator system. Specifically we re-examined recently proposed chiral mirror coupled two-resonator system from the point of view of a number non-conserving approach. In addition we proposed another heuristic model which could be realized in coupled soliton-plasmon systems. We find that as a result of the interplay between the asymmetry of the non-Hermitian coupling and initial coherence of cavity fields, a counter-intuitive population transfer can occur between the resonators. The population of an initially empty resonator can exceed that of the initially available population in the other resonator. Moreover, the excitations can exhibit collapse and revival dynamics in the presence of nonlinearity. Collapse and revival times as well as the average populations depend on the asymmetry of the non-Hermitian interaction. Finally we investigated the bipartite entanglement between the modes of the resonators using von Neumann entropy. We found that while nonlinearity induces entanglement, its dynamics and amount can be controlled by the asymmetry of the non-Hermitian  interaction. In addition, we pointed out that using Kerr-type nonlinear cavities and chiral mirrors one can simulate the double-well non-Hermitian Bose-Condensate models. Relations to $\mathcal{PT}$ symmetric lasers and Jahn-Teller systems are pointed out.

\acknowledgments
We acknowledge illuminating discussions by R.~B.~B.~ Santos, U.~G\"unther and C.~Mili\'{a}n. This work is supported by T\"UB{\.I}TAK (Grant No. 111T285). F.~A.~acknowledges the support and the hospitality of the Office of Vice President for Research and Development (VPRD) and Department of Physics of the Ko\c{c} University.

\end{document}